\documentclass[12pt,a4]{article}
\usepackage{graphicx}
\usepackage{longtable}
\usepackage{amsmath}
\usepackage[all]{xy}

\newcommand{\be}{\begin{equation}}
\newcommand{\ee}{\end{equation}}
\newcommand{\ba}{\begin{eqnarray}}
\newcommand{\ea}{\end{eqnarray}}
\newcommand{\baa}{\begin{eqnarray*}}
\newcommand{\eaa}{\end{eqnarray*}}

\begin{document}

\title{Effect of a single impurity on the local density of states in monolayer and bilayer graphene}
\author{Cristina Bena\\
{\small \it Service de Physique Th\'eorique, CEA/Saclay},
\vspace{-.1in}\\{\small \it  Orme des Merisiers, 91190 Gif-sur-Yvette CEDEX, France}}
\maketitle

\begin{abstract}
We use the T-matrix approximation to analyze the effect of a localized impurity on the local density of states in mono- and bilayer graphene. For monolayer graphene the Friedel oscillations generated by intranodal scattering obey an inverse-square law, while the internodal ones obey an inverse law. In the Fourier transform this translates into a filled circle of high intensity in the center of the Brillouin zone, and empty circular contours around its  corners. For bilayer graphene both types of oscillations obey an inverse law. 
\end{abstract}

Graphene has been studied extensively in the recent
years. Its most fascinating aspect is the existence of linearly-dispersing gapless excitations in the vicinity of the Dirac
points. This gives rise to very interesting
electronic properties such as Friedel oscillations in the local density of states at low energy which decay as $1/r^2$ \cite{falko,glazman}, instead the usual $1/r$ characteristic to two-dimensional systems
\cite{bk,steveold}. It is very important to understand well the physics of impurity scattering in
graphene, by studying for example the local density of states (LDOS) in the presence of
single impurity scattering \cite{falko,glazman,bk,others}. Comparison with experiments can provide information about graphene's fundamental physics, and about the nature of the impurities.

In this Letter we analyze the Fourier
transform of the density of states measurable by Fourier
transform scanning tunneling spectroscopy (FTSTS).
Such measurements have recently been developed for graphene
\cite{mallet}, as well as for other two-dimensional materials such as
$ErSi_2$ \cite{vonau} and the cuprates \cite{davis}.

Our first observation is that the FTSTS spectra can be used
to distinguish between monolayer and bilayer graphene.
In particular, for monolayer graphene with a localized (delta-function) impurity potential,
the Friedel oscillations generated by intranodal scattering decay
as $1/r^2$ at low STM bias, consistent with previous analysis
\cite{falko,glazman}. In the FTSTS spectra this is manifested by a filled circle of high
intensity in the center of the Brillouin zone (BZ), with a radius proportional to
the STM bias. On the other hand, the Friedel oscillations
generated by the scattering of quasiparticles between different Dirac
points decay as $1/r$. In the FTSTS spectra these oscillations are translated into
circular contours of high intensity centered around the
corners of the BZ and around sites of the reciprocal lattice. Due to the form of the underlying Hamiltonian,
the distribution of intensity on some of these circles is not rotationally invariant.

For the bilayer system, at low energy the oscillations due to both intranodal
and internodal scattering have a $1/r$ dependence, corresponding
to circular lines of high intensity close to center and the
corners of the BZ, and around the sites of the reciprocal lattice.
At higher energies the splitting of the bands for the bilayer sample is also observable in the FTSTS spectra.

We also note that the FTSTS spectra can distinguish between different types of impurities.
For example, for the case of a screened-charge impurity, the effect of internodal scattering
is greatly reduced compared to the effect of intranodal scattering. This gives a clear signature in the FTSTS spectra which can be observed in an experiment.

Our last observation is that the FTSTS spectra, besides providing
information about the band structure of graphene, can also give insight into
the underlying Hamiltonian. In particular, the shift of the decay of the Friedel oscillations from $1/r$ to $1/r^2$, and the rotational asymmetry of some of the high-intensity spots, are strongly dependent on the peculiar form of the tight-binding Hamiltonian, and cannot be deduced solely from band-structure arguments.


The tight-binding Hamiltonian for monolayer graphene is:
\be
{\cal H}=\int d^2 \vec{k} [a_{\vec{k}}^{\dagger} b_{\vec{k}} f(\vec{k})+h.c.],
\label{h0}
\ee
where the operators $a^{\dagger}$, $b^{\dagger}$ correspond to creating electrons on the sublattice
$A$ and $B$ respectively, and $f(\vec{k})=-t \sum_{j=1}^3 \exp(i \vec{k} \cdot {\vec{a}_j})$. Here
$\vec{a}_1=a(\sqrt{3} \hat{x}+\hat{y})/2$, $\vec{a}_2=a(-\sqrt{3} \hat{x}+\hat{y})/2$,
$\vec{a}_3=-a \hat{y}$, $t$ is the nearest-neighbor hopping amplitude, and
$a$ is the spacing between two adjacent carbon atoms, which we are setting to $1$.

We will use this form of the Hamiltonian when performing our numerical analysis of
the FTSTS spectra. However, it is useful to expand the Hamiltonian close to the corners of the BZ, which we also denote as nodes or ``Dirac points,'' and use the linearized form to solve the problem analytically at low energies.
The momenta of the six corners of the Brillouin zone are given by
$\vec{K}_{1,2}=[\pm 4 \pi/(3 \sqrt{3}),0]$, $\vec{K}_{3,4}=[\pm 2 \pi/(3 \sqrt{3}),2 \pi/3]$, $\vec{K}_{5,6}=[\pm 2 \pi/(3 \sqrt{3}),-2 \pi/3]$.
Close to each corner, $m$, of the BZ we
can write $f(\vec{q}+\vec{K}_{m})\approx \tilde{\phi}_{m}(\vec{q})=v_m \vec{q}\cdot\vec{J}_m$, where
$\vec{q}$ denotes the distance from the respective corner.
Also $v_{1,2}=3 t/2=v$, $v_{3,4}=v \exp(-i \pi/3)$, $v_{5,6}=v \exp(i \pi/3)$ and $\vec{J}_{1,2}=(\pm 1, -i)$, $\vec{J}_{3,4}=\vec{J}_{5,6}=(\pm 1,i)$.

The corresponding Green's function, ${\cal G}(\vec{k},\omega)$, derived from the tight-binding Hamiltonian in Eq.~(\ref{h0}) can be expanded at low energy around the six nodes (denoted $m$), and in the $2 \times 2$ (A,B) sublattice basis can be written as:
\begin{align}
{\cal G}(\vec{k},\omega)\approx G_m(\vec{k},\omega)=\frac{1}{\omega^2-|\tilde{\phi}_m(\vec{k})|^2}%
\begin{pmatrix}
\omega+i \delta & \tilde{\phi}_m({\vec{k}}) \\
\tilde{\phi}_m^*(\vec{k}) & \omega+i\delta%
\end{pmatrix}
\label{g0}
\end{align}
where $\delta$ is the quasiparticle inverse lifetime.
The Fourier transform of the linearized Green's function is given by:
\begin{align}
G_m(\vec{r},\omega)\propto \omega%
\begin{pmatrix}
H_0^{(1)}(z) & i \phi_m(\vec{r}) H_1^{(1)}(z)  \\
i \phi^{*}_m(\vec{r}) H_1^{(1)}(z) & H_0^{(1)}(z)%
\end{pmatrix}
\label{g1}
\end{align}
where $z\equiv \omega r/v $, $H_{0,1}^{(1)}(r)$ are Hankel functions, $r=|\vec{r}|$, and
$\phi_m(\vec{r})=v_m \vec{r}\cdot \vec{J}_m/(v r)$.

\begin{figure}[htbp]
\begin{center}
\includegraphics[width=4in]{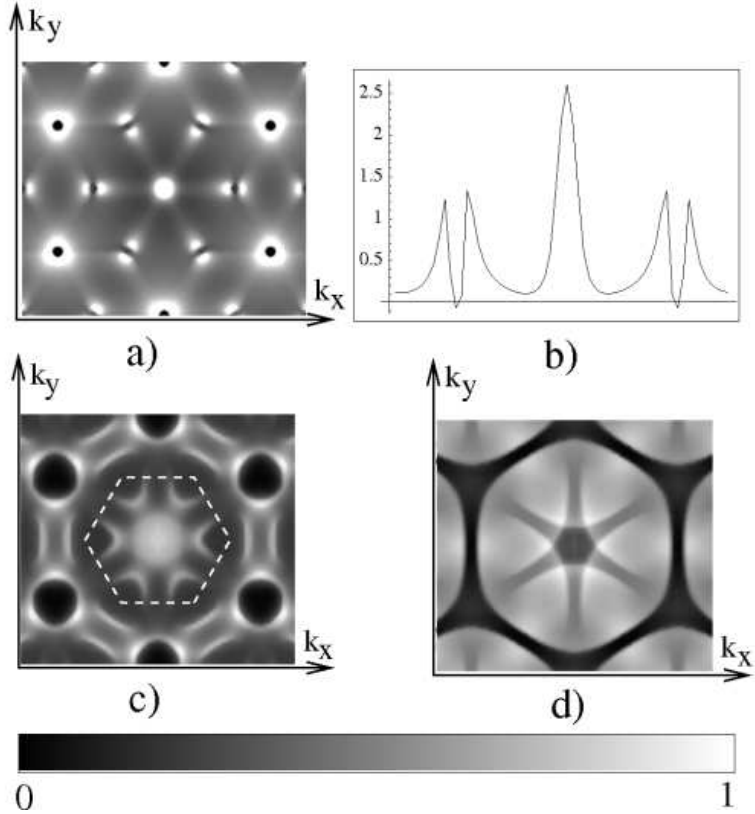}
\vspace{0.15in} \caption{\small FTSTS spectra for a monolayer graphene sample with a single delta-function impurity. Figs. 1.a), 1.c) and 1.d) correspond to energies $0.15t$, $0.6t$, and  $1.2t$ respectively, at $\delta=0.07t$.  The BZ is indicated by dashed lines. The actual lowest ($0$) and highest ($1$) values of the FTSTS intensity are different for each energy ($(-1.3,2.6)$ for $0.15t$, $(-0.8,5.9)$ for $0.6t$, and $(-6.2,7.2)$ for $1.2t$ in arbitrary units).
Fig. 1.b) shows a cross section of the FTSTS intensity as a function of $k_x$ for $k_y=0$, and for energy $0.15t$.
}
\label{fig2}
\end{center}
\end{figure}

We first focus on a delta-function impurity localized on an atom belonging to sublattice $A$. In the $(A,B)$ basis the impurity potential matrix $V$ has only one non-zero component $V_{11}=u$.
We start with a $T$-matrix study \cite{tmatrix,bk} of the full Hamiltonian presented in Eq.~(\ref{h0}), and we analyze our results numerically for various energies. The resulting FTSTS spectra (corresponding to the real part of the Fourier transform of the LDOS) are plotted in Fig.~\ref{fig2}.
There are several interesting features
that should be noted. First, there are regions of high intensity in the FTSTS spectra corresponding to intranodal quasiparticle scattering (central region) and internodal scattering (outer regions). Similar features have also been observed experimentally \cite{mallet}.
The high intensity regions that we find are point-like at zero energy, and acquire distinct features as one increases the energy
(STM bias). Thus, at low energy the central high-intensity region is a filled circle, while the outer regions are empty. Also, the rotational symmetry of the high-intensity regions located at the corners of the BZ is broken, while it is preserved for the high-intensity regions centered on sites of the reciprocal-lattice.
With increasing the energy even higher, other effects such as the changing of the shape of the equal-energy contours from circular to triangular (trigonal warping) start
playing an important role. At very high energy the FTSTS intensity map becomes
quite intricate.

We now turn to the analytical study of the dependence of the LDOS on the relative position with respect to the impurity ($\vec{r}$) at low energies. In this range, the physics is dominated by linearly dispersing quasiparticles close to the Dirac points. We find that spatial variations of the LDOS due to the impurity are given by:
\ba
\rho(\vec{r},E) \propto -{\rm{Im}}[{\cal G}(-\vec{r},E) T(E) {\cal G}(\vec{r},E)]
\approx -\sum_{m,n} {\rm{Im}}[e^{i (\vec{K}_m-\vec{K}_n)\cdot
\vec{r}}G_m(-\vec{r},E) T(E) G_n(\vec{r},E)],
\label{gr}
\ea
where $m,n$ denote the corresponding Dirac points. Here $T(E)$ is the $T$-matrix, which for a delta-function impurity is given by \cite{tmatrix}
$
T(\omega)= [I-V \int \frac{d^2 \vec{k}}{S_{BZ}}
{\cal G}(\vec{k},\omega)]^{-1}V,
$
where $I$ is the $2 \times 2$ identity matrix, and the integral over $\vec{k}$ is performed on the
BZ, whose area is $S_{BZ}=8 \pi^2/3\sqrt{3}$.

Using  Eq.~(\ref{g1}) and expanding the Hankel functions to leading order in $1/r$, we find that far from the impurity ($\omega r/v \gg 1$) the corrections to the local density of states due to scattering between the nodes $m$ and $n$ are given by:
\be
\rho_{mn}(\vec{r},\omega) \propto \frac{\omega}{r}
{\rm{Im}}\big\{t(\omega)e^{i(\vec{K}_m-\vec{K}_n)\cdot \vec{r}+2 i \omega r/v} i\big[1-
\phi_m^{*}(\vec{r}) \phi_n(\vec{r})\big]\big\}.
\ee
where $t(\omega)$ is the non-zero element of the $T$-matrix
($T_{11}$), and we used the fact that $\phi(-\vec{r})=-\phi(\vec{r})$. 

In the case of intranodal scattering ($m=n$) the above expression vanishes and the LDOS is dominated by the next leading correction $
\rho_{m}(\vec{r},\omega) \propto \sin(2 \omega r/v)/r^2$.
This is different from what usual wisdom would suggest for a two-dimensional system ($1/r$ decay) \cite{bk,steveold}, and has also been described in Refs. \cite{falko,glazman}.
We should note that the two-dimensional FT of  $\sin(2 \omega r/v)/r^2$
is roughly $\rho_m(q,\omega)\propto \pi \theta(2 \omega-q v)/2 +\arcsin(2\omega/q v) [1-\theta( 2 \omega-q v)]$. This
corresponds to a filled circle of high intensity in the FTSTS spectrum, which is consistent
with the results of our numerical analysis for the central region of high intensity.

Nevertheless, for the decay of the Fridel oscillations generated by internodal scattering
($m\ne n$), the leading order behavior is $1/r$. The FT of $\cos(2 \omega r/v)/r$ is
$\theta(q v-2\omega)/\sqrt{q^2 v^2-4 \omega^2}$, which translates into {\it empty} circles of high intensity in the
FTSTS spectra, consistent with our
numerical analysis. However, since the inverse quasiparticle lifetime $\delta$ is finite, there will be some broadening of the resonances and some weight inside the circular contours.

For some of the Fridel oscillations generated by internodal scattering, such as the ones between neighboring node pairs (e.g. $(1,3)$ ), the rotational symmetry is broken:\\ $\phi_1^{*}(\vec{r}) \phi_3(\vec{r})=e^{-i \pi/3}(x+i y)^2$.
However, for  next-to-nearest-neighbor node pairs (e.g. $(3,5)$), $\phi_3^{*}(\vec{r}) \phi_5(\vec{r}) =e^{-i 2\pi/3}$, and the oscillations are rotationally invariant. This is consistent with the results of our numerical analysis: the high-intensity regions centered on sites of the reciprocal-lattice are rotationally symmetric, while the high-intensity regions close to the corners of the BZ are not.

\begin{figure}[htbp]
\begin{center}
\includegraphics[width=4in]{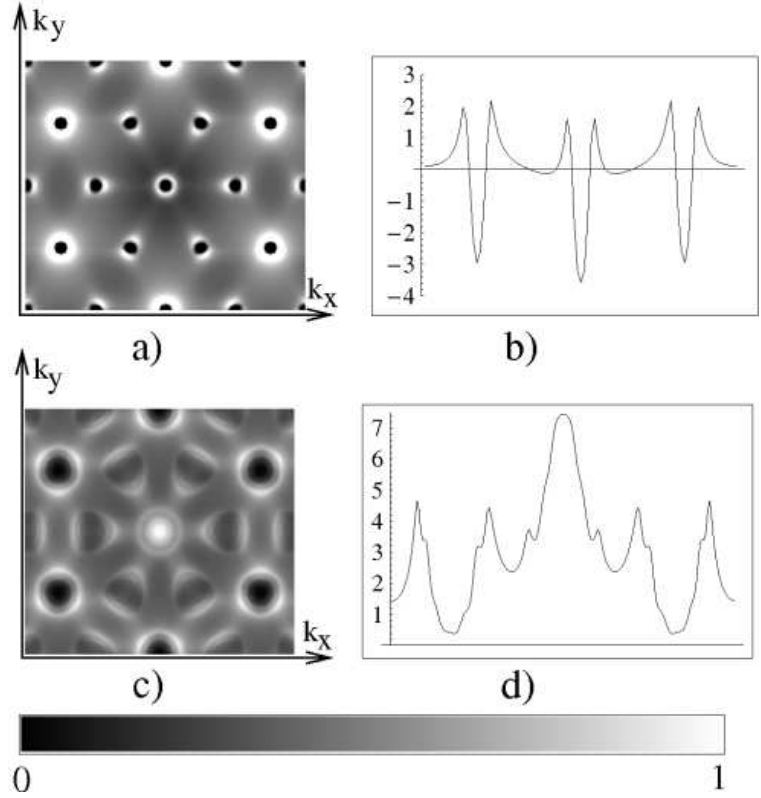}
\vspace{0.15in} \caption{\small FTSTS spectra for a bilayer sample. Fig. 2.a) and 2.c) depict the FTSTS intensity in arbitrary units at energies $0.1t$, $0.4t$, with $t_p=0.3t$, and $\delta=0.05t$. The actual lowest ($0$) and highest ($1$) values of the FTSTS intensity are $(-6.9,3.6)$ and $(-6.0,7.4)$ respectively. Figures 2.b) and 2.d) are cross-sections at $k_y=0$ of Figs. 2.a) and 2.c).}
\label{fig3}
\end{center}
\end{figure}

We now switch gears and consider the case of a single impurity in bilayer graphene. The bilayer graphene consists of two graphene layers stacked on top of each other such that
the atoms in the sublattice $A$ of the first layer occur naturally directly on top of the atoms
in the sublattice $\tilde{B}$ of the second layer \cite{bilayer1}, with a tunneling coupling of $t_p$.
We consider the case of an impurity located on the sublattice $A$.  The case of a single impurity located on a site of a different type, as well as the case of multiple impurities will be presented elsewhere. The resulting FTSTS spectra for the LDOS in the top layer are presented in Fig.~\ref{fig3}.

Note that there are similarities and discrepancies between the monolayer and bilayer cases.
Like in the monolayer case, there are areas of high intensity centered on the corners of the BZ, as well as on the sites of the reciprocal lattice. The main difference at low energy is that the central region of high intensity is an empty circle, and not a full circle (as for the monolayer case).  At high energy, we also note a doubling of the number of high intensity lines corresponding to the doubling of the number of bands.

An analytical study can be performed at low energies starting from the expansion of the
Hamiltonian around the Dirac points $m$. In the sublattice basis $(A, \tilde{B})$ this yields
\cite{bilayer2}:
\begin{align}
{\cal H}^{bilayer}_m(\vec{k})=%
\begin{pmatrix}
0 & [\tilde{\phi}_m(\vec{k})]^2 \\
[\tilde{\phi}_m^*(\vec{k})]^2 & 0%
\end{pmatrix}
\end{align}
where for simplicity we have set the effective mass of the quadratic spectrum to $1$.
The corresponding Green's function in real space is given by:
\begin{align}
G_m(\vec{r},\omega)\propto %
\begin{pmatrix}
H_0^{(1)}(z) &  -[\phi_m(\vec{r})]^2 H_2^{(1)}(z)  \\
-[\phi^*_m(\vec{r})]^2 H_2^{(1)}(z) & H_0^{(1)}(z)%
\end{pmatrix}
\label{g1b}
\end{align}
where we denoted $z=r \sqrt{|\omega|}/v$.
Starting from Eq.~(\ref{gr}), we perform a similar analysis to the case of monolayer graphene. Thus we note that at  large distances ($z \gg 1$), as opposed to the monolayer case, the leading ($1/r$) contribution
for intranodal scattering  is non-vanishing:
\be
\rho_m(\vec{r},\omega) \propto \frac{1}{r \sqrt{|\omega|}}\cos(r\sqrt{|\omega|}/v).
\ee
This is consistent with the appearance of an empty circular contour at the center of the BZ, as opposed to
the filled circle for the monolayer case.
The leading contribution to the decay of the oscillations due to internodal scattering is also $1/r$.
Note that in the monolayer case the amplitude of the Friedel oscillations corresponding to intranodal scattering is independent of energy, and the amplitude of the Friedel oscillations corresponding to internodal scattering increases as $\omega^2$, while for the bilayer case all oscillations decrease with energy as $1/\sqrt{|\omega|}$.

\begin{figure}[htbp]
\begin{center}
\includegraphics[width=4in]{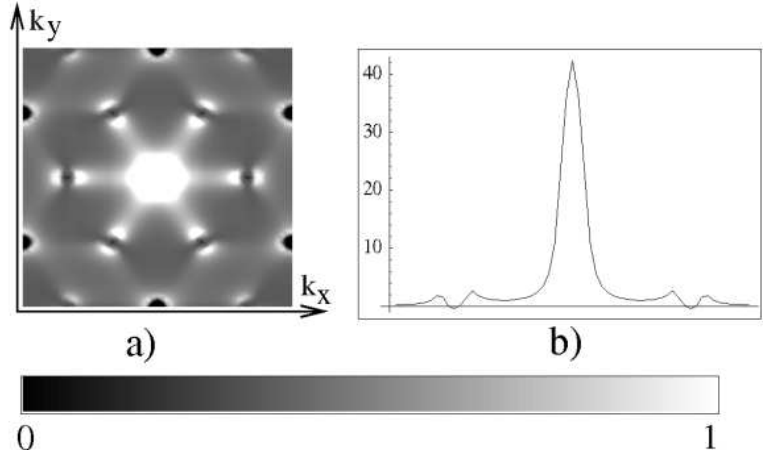}
\vspace{0.15in} \caption{\small FTSTS spectra for a graphene monolayer with a screened Coulomb impurity at energy $0.2t$, and $\delta=0.07t$. The lowest and highest values of the intensity are $(-1.7,42.3)$ in arbitrary units.
Fig. 3.b)  presents a cross-section at $k_y=0$. }
\label{fig4}
\end{center}
\end{figure}

The third system we consider is a monolayer graphene sample with a single screened-Coulomb (charged) impurity. Its impurity potential (in momentum space) has $V_{11}=V_{22}\propto 1/(q+1/\epsilon_0)$,
where $\epsilon_0\approx 4.5 a$ is the screening length \cite{coulomb}. Given the momentum
dependence of the impurity, one can no longer use the
$T$-matrix approximation, but the Born approximation.
Our results are plotted in Fig.~\ref{fig4}.
Note that the intensity of the outer areas of high intensity is
very much reduced compared to the intensity of the inner area.
This is consistent with the form of the scattering potential,
which generates less scattering between quasiparticles located on
different nodes than between quasiparticles located on the same
node. Similar qualitative differences have also been found
between the LDOS oscillations generated by the ``mirrage'' images of Coulomb-type impurities and delta-function impurities
in graphene PN junctions \cite{pn}.


To conclude, we have computed the effect of single-impurity scattering on the LDOS, in particular on the Fourier transform of the LDOS measurable experimentally by FTSTS. We have found that the FTSTS spectra in the vicinity of an impurity are a very good tool to distinguish between  monolayer and  bilayer graphene. In particular, for monolayer graphene, the Friedel oscillations due to intranodal scattering decay as $1/r^2$ and are rotationally invariant. In the FTSTS spectra they correspond to a filled high-intensity circular region at the center of the BZ. On the other hand, the Fridel oscillations due to internodal scattering decay as $1/r$. They lead to different FTSTS features, such as circular contours of high intensity centered around the corners of the BZ and on sites of the reciprocal lattice. Some of these contours display a breaking of rotational invariance.
For the bilayer case both the internodal and the intranodal Fridel oscillations decay as $1/r$.

We have also showed that the FTSTS spectra can be used to distinguish between different types of impurities, for example between a delta-function impurity and a screened Coulomb scatterer.

Last, but not least, we have noticed that, while the FTSTS spectra in the presence of an impurity can give information on the band structure, they are not fully determined by it, but also contain very important information about the specific form of the Hamiltonian. We believe that this feature is very important, and could be also used in the case of cuprates to understand the physics of high temperature superconductivity. It would be interesting for example to compare ARPES measurements and FTSTS spectra in the presence of an impurity. For graphene, this would clearly distinguish the effects of the Hamiltonian from the effects of the band structure. For other systems such as the cuprates the differences between ARPES and FTSTS will be more complex, and may provide even deeper insights into the underlying physics.

{\bf Acknowledgements} We would like to thank P. Mallet, R. Roiban, J.-N. Fuchs, M. Goerbig, G. Montambaux and F. Pi\'echon for interesting discussions. Support for this work was provided by a Marie Curie Action under FP6.


\begin{thebibliography}{99}
\bibitem{falko} V. V. Cheianov and V. I. Fal'ko, Phys. Rev. Lett. {\bf 97}, 226801 (2006).
\bibitem{glazman} E. Mariani et. al., cond-mat/0702019
\bibitem{bk} C. Bena and S. Kivelson, Phys. Rev. B {\bf 72}, 125432 (2005).
\bibitem{steveold} S. A. Kivelson et. al., Rev. Mod. Phys. {\bf 75}, 1201 (2003).
\bibitem{others} T. O. Wehling et. al., Phys. Rev. B {\bf 75}, 125425 (2007); N. M. Peres et. al., cond-mat/0705.3040; N. M. Perez, F. Guinea, and A. H. Castro Neto,
cond-mat/0512091; Phys. Rev. B {\bf 73} 125411 (2006); M. A. H. Vozmediano et. al., Phys. Rev. B {\bf 72}, 155121 (2005); T. Ando, J. Phys. Soc. Japan {\bf 75} 074716 (2006); Y. G. Pogorelov, cond-mat/0603327; Y. V. Skrypnyk and V. Loktev, Phys. Rev. B {\bf 73}, 241402(R)(2006) and Phys. Rev. B {\bf 75}, 245401 (2007); M. I. Katsnelson and A. K. Geim, arXiv:0706.2490.
\bibitem{mallet} P. Mallet et. al., Phys. Rev. B, {\bf 76}, 041403(R) (2007); G. M. Rutter et. al., Science {\bf 317} 219 (2007).
\bibitem{vonau}  F. Vonau et. al.,  Phys. Rev. Lett {\bf 95}, 176803 (2005); F. Vonau et. al. Phys. Rev. B {\bf 69}, 081305 (2004); E. Dupont-Ferrier et. al. Europhys. Lett.
{\bf 72}, 430 (2005).
\bibitem{davis} See e.g. K. McElroy, et. al., Science {\bf 309}, 1048 (2005); M. Vershinin, et. al., Science {\bf 303}, 1995 (2004); A. Fang et. al., Phys. Rev. B {\bf 70}, 214514 (2004).
\bibitem{band} P. R. Wallace, Phys. Rev. {\bf 71}, 622 (1947).
\bibitem{tmatrix} J. M. Byers, M. E. Flatt{\'e}, and D. J. Scalapino, Phys. Rev. Lett. {\bf 71}, 3363 (1993);
M. I. Salkola, A. V. Balatsky, and D. J. Scalapino, Phys. Rev. Lett. {\bf 77}, 1841 (1996);
W. Ziegler et. al., Phys. Rev. B {\bf 53}, 8704 (1996).
\bibitem{bilayer1} P. Van Mieghem, Rev. Mod. Phys {\bf 64},755 (1992); E. McCann, Phys. Rev. B {\bf 74}, 245426 (2006).
\bibitem{bilayer2}  E. McCann and V. I. Fal'ko, Phys. Rev. Lett. {\bf 96}, 086805 (2006); J. Nilsson et. al., Phys. Rev. B {\bf 73}, 214418 (2006).
\bibitem{coulomb} M. I. Katsnelson, arXiv:0706.1351.
\bibitem{pn} V. Cheianov et. al., Science {\bf 315}, 1252 (2007);

\end{thebibliography}
\end{document}